\documentclass[conference]{IEEEtran}
\usepackage[dvips]{graphicx}
\usepackage{subfigure}
\usepackage{times}
\usepackage{latexsym}
\usepackage{amsmath}
\usepackage{amssymb}
\usepackage{multirow}
\usepackage{algorithmic}
\usepackage{eucal}
\usepackage{epsfig}
\usepackage{cite}
\usepackage{longtable}
\usepackage{slashbox}
\usepackage{amsthm}
\usepackage{latexsym,bm}
\usepackage{color}
\usepackage{subfigure}
\usepackage{multicol}
\usepackage{helvet}

\theoremstyle{plain}

\theoremstyle{definition}
\theoremstyle{definition}

\begin{document}

\title{{ Performance of a Multiple-Access DCSK-CC System over Nakagami-$m$ Fading Channels}}
\author{\IEEEauthorblockN{Yi Fang and Lin Wang}
\IEEEauthorblockA{Department of Communication Engineering
\\Xiamen University, Xiamen, Fujian 361005, China
%\\ ${^\S}$Department of Electronic Engineering City University of Hong Kong, Hong Kong SAR, China
\\Email: wanglin@xmu.edu.cn, fangyi1986812@gmail.com}
\and
\IEEEauthorblockN{Guanrong Chen}
\IEEEauthorblockA{Department of Electronic Engineering\\
City University of Hong Kong, Hong Kong SAR, China\\
Email: eegchen@cityu.edu.hk}}

\maketitle

\begin{abstract}
In this paper\footnote{This work was supported by China NSF under Grants 61001073, 61102134 and 61271241.}, we propose a novel cooperative scheme to enhance the performance of multiple-access (MA) differential-chaos-shift-keying (DCSK) systems. We provide the bit-error-rate (BER) performance and throughput analyses for the new system with a decode-and-forward (DF) protocol over Nakagami-$m$ fading channels. Our simulated results not only show that this system significantly improves the BER performance as compared to the existing DCSK non-cooperative (DCSK-NC) system and the multiple-input multiple-output DCSK (MIMO-DCSK) system, but also verify the theoretical analyses. Furthermore, we show that the throughput of this system approximately equals that of the DCSK-NC system, both of which have prominent improvements over the MIMO-DCSK system. We thus believe that the proposed system can be a good framework for chaos-modulation-based wireless communications.
\end{abstract}

\section{Introduction}
In wireless communication applications, such as wireless personal area networks (WPAN) and sensor networks (WSN), multipath fading is a major factor that deteriorates the quality of information transmission. As a spread-spectrum modulation, differential chaos shift keying (DCSK) offers a promising solution to mitigate the effect of fading in such systems \cite{Kolumbam96}.

The DCSK communication system can be easily implemented in hardware since it can work without synchronization nor channel estimation but requiring only frame and symbol rate samplings, which makes it very promising in WPAN and WSN applications \cite{5383640}. In recent years, some variants of the DCSK modulation technique have been proposed \cite{972858,journals/ijbc/XuWK11}. Aiming to further overcome the signal fading arising from multipath propagation, cooperative diversity has been applied to the conventional DCSK system to construct a two-user cooperative DCSK system \cite{5537139,5629387}.

Another desirable application of DCSK is to be combined with multiple-access (MA) techniques. Recently, a large amount of research work have been devoted to MA-DCSK systems \cite{974883,1015003}. In particular, the Walsh code has been used to ensure the orthogonality of DCSK channels \cite{1015003}, so that the interference among different users can be avoided. In \cite{Fang2012TCAS}, the multiply-antenna relay was adopted in the MA-DCSK system (i.e., MIMO-DCSK system) to increase the robustness against signal fading. However, to the best of our knowledge, in all existing MA-DCSK systems, cooperative communication (CC) technique has never been applied, for which the cooperative method developed in \cite{5537139,5629387} is not applicable.

In this paper, we propose a novel cooperative scheme for the MA-DCSK systems, forming a MA-DCSK-CC system, to improve the performance of communications. We analyze the bit-error-rate (BER) performance and the throughput of the new system under a decode-and-forward (DF) relaying protocol over Nakagami-$m$ fading channels. Both theoretical analyses and computer simulations demonstrate that the proposed system has significant performance improvement in comparison with the DCSK non-cooperative (DCSK-NC) system and the MIMO-DCSK system. Moreover, we show that the throughput of the proposed system is almost the same as that of the DCSK-NC system, superior to that of the MIMO-DCSK system.

\section{System Model}
\label{sect:system}

\subsection{Overview of MA-DCSK System} \label{sect:multi_DCSK}
For an $N$-user MA-DCSK system, the orthogonal Walsh code sequences are adopted to eliminate interference among users (i.e., there is no interference among users if Walsh code is used) \cite{1015003}. In such a system, the $2^n$-order Walsh code is defined to accommodate $N$ users, where
\begin{equation}
W_{2^n}= \left (
\begin{array}{rrl}
 W_{2^{n-1}} &  W_{2^{n-1}} \cr
 W_{2^{n-1}} & -W_{2^{n-1}}
\end{array}
\right),  \quad n=1,2,...
\label{eq:Walsh}
\end{equation}
in which $2^n =2N$ and $W_{2^0}= W_1 = 1$.

Each DCSK modulated signal includes $2N$ sub-segments. Let $\beta$ denote the length of each carrier segment (i.e., sub-spreading factor). Then, the global spreading factor is kept at $2N\beta$. The $l$th transmitted signal of the $K$th user is given by
\begin{equation}
s_{K, b_l} = \sum\nolimits_{j=0}^{2N-1} w_{2K-b_l,j}\,c\left(t-j \frac{T} {2N}\right), \quad 0 < t < T
\label{eq:multi_signal}
\end{equation}
where $b_l = \{0, 1\}$ represents the $l$th transmitted symbol, $w_{i,j}$ represents the $(i, j)$th element of the $2N$-order Walsh code, and $c(t)$ represents the chaotic carrier.

At the receiver, we assume perfect timing synchronization and utilize the generalized maximum likelihood (GML) detector \cite{1329074} to demodulate the received signals.

\subsection{MA-DCSK-CC System} \label{sect:DCSK-CC}
\begin{figure}[tbp]%[t]
\center
\includegraphics[width=3.5in,height=1.1in]{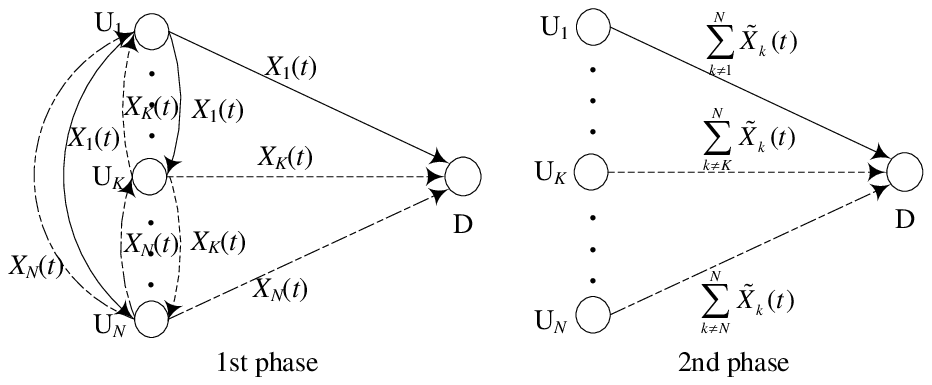}
\vspace{-0.6cm}
\caption{Basic model of the $N$-user MA-DCSK-CC system.}
\label{fig:Fig.1}
\vspace{-0.2cm}
\end{figure}

\begin{figure}[tbp]%[t]
\center
\includegraphics[width=3.5in,height=1.1in]{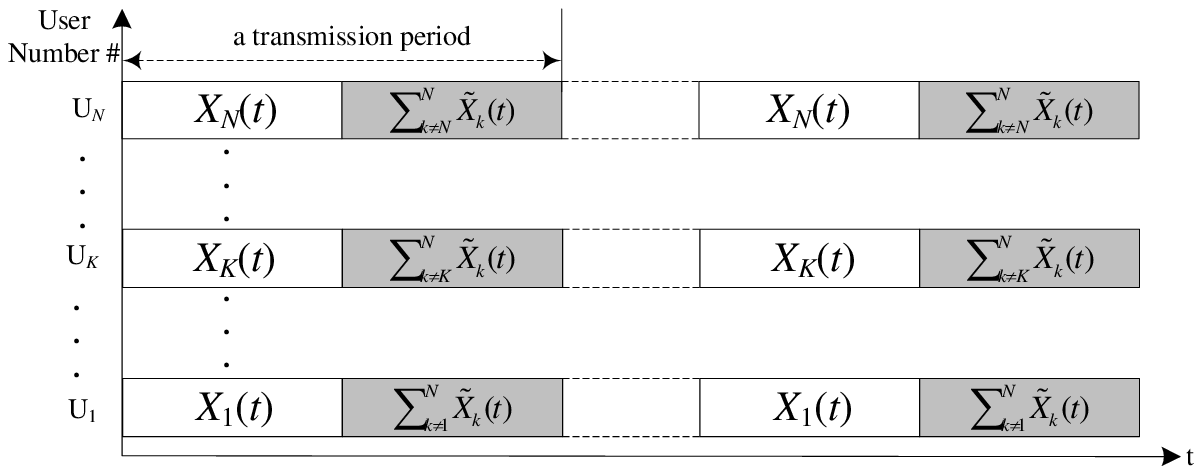}
\vspace{-0.7cm}
\caption{Cooperative scheme of the $N$-user MA-DCSK-CC system.}
\label{fig:Fig.2}
\vspace{-0.4cm}
\end{figure}

The full-duplex system model is shown in Fig.~\ref{fig:Fig.1}. Referring to this figure, the transmission period is divided into two phases, namely $1$st phase and $2$nd phase. In the $1$st phase, the $K$th ($K=1 \sim N$) user $ U_K$ broadcasts its message $X_K(t)$ to other terminals. In the $2$nd phase, $U_K$ helps the remainder $N-1$ to forward their messages. Using the DF protocol, $U_K$ transmits the reconstructed signals, i.e., $F (\widetilde{X}_1(t),\cdots,\widetilde{X}_{K-1}(t),\widetilde{X}_{K+1}(t),\cdots,
\widetilde{X}_{N}(t))= \sum_{k \neq K}^{N} \widetilde{X}_k(t)$ ($\widetilde{X}_k(t)$ is the reconstructed signal of $X_k(t)$), to the destination
if it decodes correctly; otherwise, it remains idling. Based on the above description, the cooperative scheme of the system is illustrated in
Fig.~\ref{fig:Fig.2}. According to such a scheme, the received signals are expressed as
%Fig.1
%Fig.2
\begin{eqnarray}
 r_K^1 (t) &=& \sum\nolimits_{k \neq K}^{N} H_{k, K} \otimes X_k (t)  + Z_K^1 (t)\\
 r_D^1 (t) &=& \sum\nolimits_{k = 1}^{N} H_{k, D} \otimes X_k (t) + Z_D^1 (t) \\
 r_D^2 (t) &=& \sum_{K = 1}^{N} \left(H_{K,D} \otimes \sum\nolimits_{k \neq K}^{N} \widetilde{X}_k(t) \right) + Z_D^2 (t)
\label{eq:DCSK-CC-2-receiver}
\end{eqnarray}
Here, the superscripts ``1'' and ``2'' denote the $1$st phase and the $2$nd phase, respectively; $[K, k] = 1\sim N$; $\otimes$ denotes convolution; $r_K^1 (t)$ represents the received signals at $U_K$ in the $1$st phase while $r_D^i (t)$ ($i=1,2$) represents the received signals at destination in the $i$th phase; $Z_R^{i}(t)$ and $Z_D^{i}(t)$ are white Gaussian noise random variables (RV) with zero mean and variance $N_0/2$; $H_{k,K}$, $H_{k,D}$, and $H_{K,D}$ are the channel impulse responses of links $U_k \to U_K$, $U_k \to {\rm D}$, and $U_K \to {\rm D}$, respectively, which can be represented as
\begin{equation}
H = \sum_{l=1}\nolimits^{L} \alpha_l \delta (t - \tau_l)
\label{eq:CIR}
\end{equation}
where $\alpha_l$ and $\tau_l$ are the attenuation and time delay of the $l$th path, respectively, and $L$ is the number of multipaths.
Moreover, the attenuation $\alpha_l$ follows a Nakagami-$m_l$ distribution, with the probability density function (PDF) given by
\begin{equation}
f_{\mid \alpha_l\mid} (x)= \frac{2} {\Gamma (m_l)} \left( \frac{m_l} {\Omega_l} \right)^{m_l} x^{2 m_l - 1}
\exp \left( - \frac{m_l} {\Omega_l} x^2 \right)
\label{eq:Nakagami_PDF}
\end{equation}
where $m_l \geq 0.5$ denotes the fading factor of the $l$th path, $\Omega_l$  equals $E(\alpha_l^2)$, and $\Gamma(\cdot)$ is the Gamma function. For ease of analysis, assume that the channels possess an uniform scale parameter, i.e., $\Omega_l$ ($\Omega_1=\Omega_2=\cdots=\Omega_L=\Omega$) and $m_l$ ($m_1=m_2=\cdots=m_L=m$) are kept constant for all path-components. Also, assume that the receiver can capture all paths power, i.e., $\sum_{l=1}^{L} \Omega_l= \sum_{l=1}^{L} E(\alpha_l^2)=1$.

\section{Performance Analyses} \label{sect:Performance}
In this section, the BER performance and throughput of the proposed MA-DCSK-CC system are analyzed over a Nakagami-$m$ fading channel. To simplify the analysis, assuming that the energy per bit of each user is constant, denoted by $E_b$, and the energy is allocated uniformly to every transmission phase. Moreover, we assume that the time delay $\tau_l$ is much shorter than the bit duration ($2N \beta T_s \gg \tau_l$, where $T_s$ is the sampling period) such that the inter-symbol interference can be ignored. Also, the equal-gain combiner (EGC) is employed at the receiver for easy implementation.

\subsection{BER Performance} \label{sect:BER-analysis}
Here, we derive the BER expression of $U_K$ in the $N$-user MA-DCSK system, since all users in such a scenario have the same error performance. Let $d_{SD}$, $d_{SR}$, $d_{RD}$ denote the distance from source to destination (S $\to$ D), source to relay (S $\to$ R) and relay to destination (R $\to$ D) links, respectively; hence, $d_{SD}:d_{SR}:d_{RD}$ represents the geometric positions of all terminals. The path loss of a (S $\to$ D) link is defined by $PL_{SD}=1/d_{SD}^2$.

The total BER under the DF protocol is described as
\begin{equation}
%%%%%%\vspace{-5mm}
{\rm BER}_{\rm CC-DF}  = {\rm BER}_{SR} \cdot {\rm BER}_{SD} + (1 - {\rm BER}_{SR}) \cdot {\rm BER}_{D}
\label{eq:CC-DF-BER}
\end{equation}
where ${\rm BER}_{SR}$, ${\rm BER}_{SD}$, and ${\rm BER}_D$ denote the BERs at the relay receiver with the signal from source, at the destination receiver with the signal from source, and at the destination receiver with all signals from source and relays, respectively.

Let $\gamma$ denote the received signal-to-noise ratio (SNR) of a link. Then, the BER of such a link can be obtained by averaging the conditional BER of $\gamma$, as follows:
\begin{equation}
{\rm BER} =  \int_{0}^{\infty} {\rm BER}(\gamma) f(\gamma) {\rm d} \gamma
%\vspace{-1cm}
\label{eq:link-BER-expresion}
\end{equation}
where ${\rm BER}(\gamma)$ and $f(\gamma)$ are the conditional BER and PDF of the received SNR. Specifically, ${\rm BER}(\gamma)$ is given by \cite{1362943}
\begin{equation}
{\rm BER} (\gamma) = \frac{1} {2}\, \text{erfc} \left( \left[\frac{4} {\gamma}
\left(1 + \frac{\beta} {2 \gamma}\right)^{- \frac{1} {2}} \right] \right)
\label{eq:condi-BER}
\end{equation}
To obtain the explicit expressions of ${\rm BER}_{SR}$, ${\rm BER}_{SD}$ and ${\rm BER}_D$. We firstly need to deduce the PDF of received SNRs of the three links: $f (\gamma_{SR})$, $f (\gamma_{SD})$ and $f (\gamma_{RD})$.

In the $1$st phase, the received SNR of link S $\to {\rm R}_k$ (i.e., $U_K \to U_k$, $k \neq K$) can be written as
\begin{equation}
\gamma_{SR_k} = \sum\nolimits_{l=1}^{L}  \gamma_l = (E_b/N_0) (2 d_{SR_k}^2) \sum\nolimits_{l=1}^{L} \alpha_l^2
\label{eq:SR-rece_SNR}
\end{equation}
where $E_b$ is the bit energy (the energy consumed in the $1$st phase is $E_b /2$), and $\gamma_l$ is the instantaneous SNR of the $l$th path. It is well known that the square of a Nakagami-$m$-distributed RV follows a gamma distribution, denoted as $G (m, \Omega / m)$, with the PDF as follows:
\begin{equation}
f (x) = \frac{x^{m-1} e^{-x/(\Omega / m)}} {(\Omega / m)^m \Gamma(m)}, \quad x>0
\label{eq:Gamma_PDF}
\end{equation}
Based on the Nakagami-$m_l$-distributed RV $\alpha_l$, we obtain the PDF of $\gamma_l$, as
$\gamma_l = \frac{E_b} {2 N_0 d_{SR_k}^2} \alpha_l^2 \sim G (m_l, \frac {E_b \Omega_l} {2 N_0 m_l d_{SR_k}^2} )$. Hence, $\gamma_{SR}$ is further derived as \cite{Fang2012TCAS}
\begin{equation}
\begin{array}{rl}
\gamma_{SR_k} \sim &  \hspace{-2mm} G \left(\sum_{l=1}^{L} m_l, \frac{ E_b \sum_{l=1}^{L} \Omega_l} {2 N_0 d_{SR_k}^2 \sum_{l=1}^{L} m_l} \right)
\vspace{1mm}\\
= &  \hspace{-2mm} G \left(mL, \frac{E_b/N_0} {2 m L d_{S R_k}^2} \right)\hspace{-0.5mm} = G (x_1, y_1)
 \end{array}
\label{eq:SR-PDF}
\end{equation}
Here, $m_l=m$, $d_{S R_k}^2=d_{SR}^2$, $x_1$ and $y_1$ are used as short-hand notations for $mL$ and $(E_b/N_0)/(2mL d_{SR_k}^2)$, respectively. As the $N-1$ relays decode the received signal independently, the equivalent received SNR of link S $\to$ R equals $\gamma_{SR_k}$, namely, $\gamma_{SR}=\gamma_{SR_k}$.

Likewise, the received SNR of link S $\to$ D is expressed by
\begin{equation}
\hspace{-1mm}\gamma_{S D} \hspace{-0.5mm} = \hspace{-0.5mm} \frac{E_b} {2 N_0 d_{S D}^2} \sum_{l=1}^{L} \alpha_l^2 \vspace{1mm}\\
\sim \hspace{-0.5mm} G \left(mL, \frac{E_b/N_0} {2 m L d_{S D}^2} \right)\hspace{-0.5mm} = G (x_2, y_2)
% \end{array}
\label{eq:SD-PDF}
\end{equation}
where $x_2$ and $y_2$ are short-hand notations for $mL$ and $(E_b/N_0)/(2 m L d_{SD}^2 )$, respectively.

Moreover, in the $2$nd phase, one can get the distribution of the received SNR of link ${\rm R}_k \to$ D as
$\gamma_{R_k D} \sim G (mL, \frac{E_b/N_0} {2 (N-1) m L d_{R_k D}^2} )$,
where $d_{R_k D}^2=d_{RD}^2$. Adopting the EGC method, $\gamma_{RD}$ is readily obtained through combining all $N-1$ received SNR at the destination receiver, resulting in
\begin{equation}
\begin{array}{rl}
\hspace{-4mm} \gamma_{R D} = \hspace{-1mm} \sum_{k\neq K}^{N} \gamma_{R_k D}
\sim & \hspace{-2mm}  G \left((N-1)mL, \frac{E_b/N_0} {2 (N-1) m L d_{R D}^2} \right) \vspace{1mm}\\
= & \hspace{-2mm} G (x_3, y_3)
 \end{array}
\label{eq:RD-PDF}
\end{equation}
where $x_3$ and $y_3$ are short-hand notations for $(N-1)mL$ and $(E_b/N_0)/(2(N-1)mL d_{R D}^2)$, respectively.
Afterwards, the total received SNR at destination with all signals from source and relays can be represented as $\gamma_D = \gamma_{SD} +\gamma_{RD}$. Thus, the corresponding PDF can be formulated by \cite{Moschopoulos1985}
\begin{equation}
\hspace{-0.5mm} f (\gamma_D)  = \left\{\begin{array}{rl}
\frac{\gamma_D^{x_2 + x_3 - 1} e^{- \gamma_D / y_2}} {y_2^{x_2+x_3} \Gamma (x_2+x_3)} &\mbox{ if $y_2 = y_3$}\vspace{1mm}\\
C {\sum_{i=0}^{\infty}} \left( \frac{\xi_i \gamma_D^{\rho + i - 1} e^{- \gamma_D / y_0}} { \Gamma (\rho + i) y_0^{\rho + 1}} \right) &\mbox{ if $y_2 \neq y_3$}
\end{array} \right.
\label{eq:SDRD-PDF}
\end{equation}
Here, the parameters for the case of $y_2 \neq y_3$ are subjected to
\begin{equation}
\left\{
\begin{aligned}
& C = \prod\nolimits_{k=2}^{3} (y_0 / y_k )^{x_k}\\
&\xi _{i + 1} = \frac{1} {i + 1} \sum\nolimits_{j=1}^{i + 1} j z_j \xi_{i+1-j} &\mbox{ $i = 0,1,2,\cdots$}\\
& z_j = \sum\nolimits_{k=2}^{3} x_k ( 1 - y_0 / y_k)^j/j &\mbox{ $j =1,2,\cdots$}\\
& \rho = \sum\nolimits_{k=2}^{3} x_k > 0\\
& y_0 = \min\nolimits_{k=2}^{3} (y_k)
\end{aligned}
\right.
\label{eq:indepsum-Gamma-parameters}
\end{equation}
where $\xi_0 = 1$.

Substituting \eqref{eq:condi-BER} and \eqref{eq:SDRD-PDF} into \eqref{eq:link-BER-expresion} yields
\begin{equation}
{\rm BER}_{D} =  \int_{0}^{\infty} {\rm BER}(\gamma_D) f(\gamma_D) {\rm d} \gamma_D
%\vspace{-1cm}
\label{eq:D-BER}
\end{equation}
Similarly, with the help of \eqref{eq:condi-BER} (the conditional BER expressions of different links are identical), ${\rm BER}_{SR}$ and ${\rm BER}_{SD}$ can be calculated by substituting \eqref{eq:SR-PDF} and \eqref{eq:SD-PDF} into \eqref{eq:link-BER-expresion}, respectively, giving
\begin{equation}
{\rm BER}_{SR} =  \int_{0}^{\infty} {\rm BER}(\gamma_{SR}) f(\gamma_{SR}) {\rm d} \gamma_{SR}
%\vspace{-1cm}
\label{eq:SR-BER}
\end{equation}
\begin{equation}
{\rm BER}_{SD} =  \int_{0}^{\infty} {\rm BER}(\gamma_{SD}) f(\gamma_{SD}) {\rm d} \gamma_{SD}
%\vspace{-1cm}
\label{eq:SD-BER}
\end{equation}
Finally, by combining \eqref{eq:D-BER}, \eqref{eq:SR-BER}, \eqref{eq:SD-BER}, and \eqref{eq:CC-DF-BER}, the BER of the proposed system, i.e., ${\rm BER}_{\rm CC-DF}$, is formulated.

\subsection{Throughput} \label{sect:throughput}
Assume that the normalized throughput is defined as the average number of successfully received bits/symbols after error detection in each transmission period (as in \cite{1246003}). Consequently, the throughput $\eta$ is a decreasing function of the BER. Based on such a definition, the normalized throughput of the proposed system is given by
\begin{equation}
\eta_{\rm CC} =  1 - {\rm BER}_{\rm CC-DF}
%\vspace{-1cm}
\label{eq:CC-Throu}
\end{equation}
Also, the normalized throughput of the MA-DCSK-NC system \cite{1015003} and MIMO-DCSK system \cite{Fang2012TCAS} are as follows:
\begin{equation}
\eta_{\rm NC} =  1 - {\rm BER}_{\rm NC-DF}
%\vspace{-1cm}
\label{eq:NC-Throu}
\end{equation}
\begin{equation}
\eta_{\rm MIMO} =  \frac{N-1} {N} (1 - {\rm BER}_{\rm MIMO-DF})
%\vspace{-1cm}
\label{eq:MIMO-Throu}
\end{equation}
where the BER expressions of the DCSK-NC system, i.e., ${\rm BER}_{\rm NC-DF}$, and the MIMO-DCSK system, i.e., ${\rm BER}_{\rm MIMO-DF}$,
are shown in \cite{1015003} and \cite{Fang2012TCAS}, respectively. One should note that $(N-1)/N$ in \eqref{eq:MIMO-Throu} is the normalized
factor because one of the $N$ users never transmits any message in the MIMO-DCSK system. Thus, \eqref{eq:MIMO-Throu} reduces to
$\eta_{\rm MIMO} \approx  1 - {\rm BER}_{\rm MIMO-DF}$ if $N$ is enough large.
%%%%%%%%%%%%%%%%%%%%%%%%%%

\section{Simulation Results} \label{sect:SIMU}
Here, we provide some simulation results for the proposed system. All the simulations were performed in $4$-user MA systems ($N=4$) over Nakagami fading channels with parameters $m=2$, $L=2$, $(\tau_1, \tau_2) =(0,T_s)$, and $d_{SD}: d_{SR}: d_{RD}=1:1:1$. The sub-spreading factor $\beta=32$, such that the bit duration is much larger than the time delay, i.e., $2N \beta T_s \gg \tau_l$. As assumed, the transmission energy per bit of each user is kept constant.

\begin{figure}[tbp]
\center
\includegraphics[width=2.8in,height=1.85in]{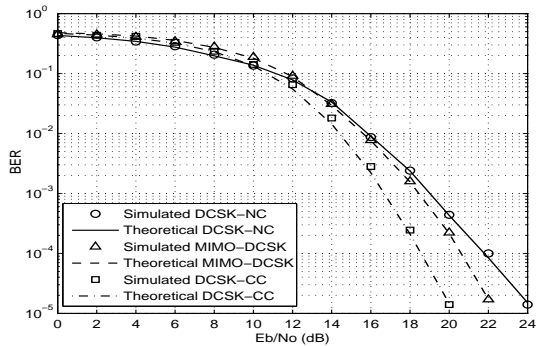}
\vspace{-0.3cm}
\caption{BER performance of the three MA-DCSK systems.
}
\label{fig:Fig.3}
\end{figure}

\begin{figure}[tbp]
\center
\includegraphics[width=2.88in,height=1.85in]{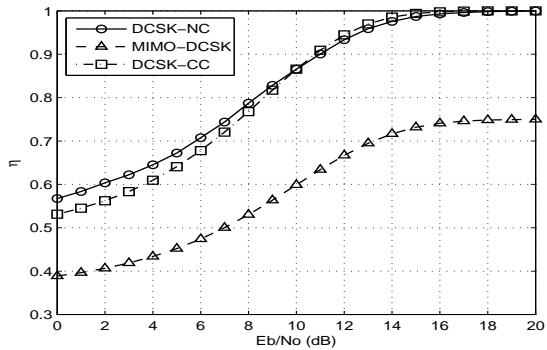}
\vspace{-0.3cm}
\caption{Normalized throughput of the three MA-DCSK systems.}
\label{fig:Fig.4}
\end{figure}

For comparison, we also consider two existing MA-DCSK systems, namely the DCSK-NC system and the MIMO-DCSK system. Fig.~\ref{fig:Fig.3} and Fig.~\ref{fig:Fig.4} present the BER performance and throughput of the these three systems. One can observe from Fig.~\ref{fig:Fig.3} that the DCSK-CC system and the DCSK-NC system are the best-performing one and the worst-performing one, respectively. At a BER of $2 \times 10^{-5}$, one can also observe that the DCSK-CC system has a gain about $0.2$~dB and $0.4$~dB over the MIMO-DCSK and DCSK-NC systems, respectively. Moreover, the theoretical BER curves are highly consistent with the simulation curves, demonstrating the correctness of the analytic results.

Referring to Fig. 4, the MIMO-DCSK system has the lowest throughput since one of the four users only helps the others to transmit messages rather than to send messages by itself. As can be seen, the proposed system overcomes this weakness and obtains a significant throughput gain as compared to the MIMO-DCSK system. Furthermore, the proposed system exhibits almost the same throughput as the DCSK-NC system, i.e., the throughput of the proposed system is slightly lower than that of the DCSK-NC system in the low-SNR region but is slightly higher when $E_b/N_0$ exceeds $10$~dB.
%Fig. 3
%Fig.4

\begin{figure}[tbp]
\center
\includegraphics[width=2.8in,height=1.85in]{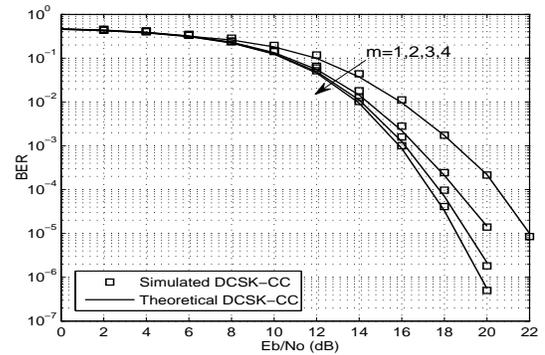}
\vspace{-0.3cm}
\caption{BER performance of the MA-DCSK-CC system with different fading depths.}
\label{fig:Fig.5}
\vspace{-0.2cm}
\end{figure}

In Fig.~\ref{fig:Fig.5}, we compare the BER curves of the proposed system for different fading depths. The parameters used remain unchanged, except that the
fading factor $m$ is varied from $1$ to $4$. As expected, one can see that the BER performance is improving as $m$ is increasing. However, the rate of
improvement is reduced with the increase of the fading factor. For instance, at a BER of $10^{-5}$, the improved gain are $2$~dB, $1$~dB, and $0.5$~dB,
respectively, as $m$ increases from $1$ to $2$, $2$ to $3$ and $3$ to $4$.
%Fig.5

\section{Conclusions} \label{sect:Conclusions}
In this paper, we have proposed, analyzed and simulated a MA-DCSK-CC system, which can enhance the anti-fading capability in wireless networks. Specifically, we have analyzed the BER performance and throughput of the new system with the DF protocol over Nakagami-$m$ fading channels, which agree with the simulation results almost perfectly. Comparing with the existing DCSK-NC and MIMO-DCSK systems, the proposed system not only shows excellent BER performance but also possesses satisfactory throughput. Therefore, among the three systems, the new DCSK-CC system can provide the best possible balance between the BER performance and the throughput. We believe that the proposed system is very suitable for low-power and low-cost WPAN applications.

\iffalse

\fi
%\bibliographystyle{IEEEtran}
%\bibliography{IEEEabrv,bib/TCAS_refs_2011}

\end{document}